# Vortex glass-liquid transition and activated flux motion in an epitaxial, superconducting NdFeAs(O,F) thin film

J. Hänisch[1*], K. Iida[2,3], T. Ohmura[3], T. Matsumoto[2], T. Hatano[2,3], M. Langer[1], S. Kauffmann-Weiss[1], H. Ikuta[2,3], B. Holzapfel[1]

[1]Institute for Technical Physics, Karlsruhe Institute of Technology, 76344 Eggenstein-Leopoldshafen, Germany

[2]Department of Materials Physics, Nagoya University, Furo-cho, Chikusa-ku, Nagoya 464-8603, Japan

[3]Department of Crystalline Materials Science, Nagoya University, Furo-cho, Chikusa-ku, Nagoya 464-8603, Japan

*email: jens.haenisch@kit.edu

## Abstract

An epitaxial NdFeAs(O,F) thin film of 90 nm thickness grown by molecular beam epitaxy on MgO single crystal with $T_c$ = 44.2 K has been investigated regarding a possible vortex glass-liquid transition. The voltage-current characteristics show excellent scalability according to the vortex-glass model with a static critical exponent ν of around 1.35 and a temperature-dependent dynamic exponent $z$ increasing from 7.8 to 9.0 for the investigated temperature range. The large and non-constant $z$ values are discussed in the frame of 3D vortex glass, thermally activated flux motion, and inhomogeneity broadening.

**Introduction**

Since Koch *et al*. [1] introduced the theory of a glassy vortex state in high-temperature superconductors as early as 1989, the characteristic scaling of the resistivity $\rho$ as well as the voltage-current, $V(I)$, respectively electric field-current density, $E(J)$, curves near its transition to the vortex liquid state have been found in numerous high-temperature cuprate superconductors. Even though the vortex liquid phase is expected primarily for superconductors with large Ginzburg numbers $Gi$, corresponding to the presence of strong thermal fluctuations, the glass-liquid transition (GLT) has also been reported in superconductors with small $Gi$, e.g. in Nb and In films, alloys, and $MgB_2$.

For the Fe-based superconductors(FBS), the GLT has also been investigated in all relevant material classes, such as 11 compounds (e.g. thin films of Fe(Se,Te) [2]), 111 compounds (e.g. LiFeAs [3]), 122 compounds (e.g. Co- doped $BaFe_2As_2$ (Ba122) [4]), and the Sm1111 compounds ($SmFeAsO_{1-x}$ single crystals [5] and SmFeAs(O,F) polycrystalline bulk samples [6]). On the other hand, for most other $Ln$1111 compounds ($Ln$ lanthanoid), e.g. La1111 and Gd1111 (as well as Sr-based 122 compounds), literature data on a possible GLT seem not available, and for Nd1111 compounds, the available data are rare and inconclusive. Liu *et al*. [7] found a signature of a GLT for $NdFeAsO_{1-x}$ polycrystals but no sign for such a transition in NdFeAs(O,F). Xie *et al*. [8] showed simple resistivity scaling for literature data on $NdFeAsO_{0.7}F_{0.3}$ polycrystals. So far, proper GLT scaling of $V(I)$ curves in combination with resistivity data is missing for NdFeAs(O,F), especially for single crystals or epitaxial thin films. This study will present such data on a thin-film sample for the first time in order to gain deeper understanding of the vortex behavior near the irreversibility line in NdFeAs(O,F) in particular and in Ln1111 materials and Fe-based superconductors in general.

The class of Ln1111 compounds is not only interesting for basic investigations of FBS but potentially also for high-field and detector applications due to their relatively high $T_c$ values of up to 56 K (highest among FBS), high upper critical and irreversibility fields as well as low electronic anisotropies at low temperatures, even though the preparation of films, tapes and wires of these materials is not as straightforward as for 11 and 122 compounds. In order to successfully increase the irreversibility fields and current carrying capabilities, it is helpful to know the behavior of the vortices near the irreversibility line.

**Sample Preparation and electrical measurements**

Epitaxial NdFeAs(O,F) thin films of around 90 nm thickness were fabricated on MgO(001) single crystals by molecular beam epitaxy (MBE). For fluorine doping, the NdOF over-layer technique has been employed in which the mother compound NdFeAsO is deposited, followed by a NdOF over-layer. Both NdFeAsO and NdOF were deposited at 800 °C. More details about sample preparation can be found in Ref. [9].

For the electrical transport measurements, a microbridge of 30 µm width and 1 mm length was structured by laser cutting. The $V(I)$ characteristics as well as the temperature dependence of the resistance, $R(T)$, were measured in four-point geometry in a Quantum Design 14 T PPMS with Keithley current source 2460 and Nanovoltmeter 2482A. The current and voltage leads were attached at the contact pads via Pogo pins. The film investigated had an onset $T_c^{on}$ of 44.2 K, which is determined as the extrapolation of largest slope in $R(T)$ towards the tempera-ture dependent normal state resistance $R_n(T)$, corresponding to around 90% of $R_n$, Fig. 1(a). The upper critical field $B_{c2}(T)$ was determined with the same 90% $R_n$ criterion. The $V(I)$ characteristics were measured up to a voltage level of 0.1 mV for constant magnetic fields ($B||c$) between 1 and 13 T (step size 2 T) at temperature steps of 0.4 K between 4 K and 35 K. The $\ln V(\ln I)$ curves were fitted with first and second order polynomials with different fit procedures to extract glass-liquid temperature $T_g$ (disappearance of quadratic term) and the $n$ value ($n = d\ln V/d\ln I$) near the critical current density $J_c$, which itself was determined using an electric field criterion of 1 μV/cm.

**Results and Discussion**

The resistive transition to the superconducting state of the film, Fig. 1(a), shows the typical broadening with increasing magnetic fields ($B||c$) due to thermally activated flux motion. The activation energy $U_0$ for flux motion is usually determined by linear fits to the relevant region in the respective Arrhenius plots, Fig. 1(b). This analysis is, however, only valid for activation energies which are either constant or linearly dependent on temperature. In those cases, the derivative $U' = - d\ln R/dT^{-1}$ is temperature-independent. As shown in Fig. 2(a), this is not the case for the investigated sample. To evaluate this further, we follow here the analysis by Zhang *et al.* [6] based on the early work of Palstra *et al.* [10]: For activated flux motion, the resistance is approximated by

$$R(T,B) = (2R_0 U(B,T)/kT) \exp(-U(B,T)/kT) \tag{1}$$

where the prefactor $2R_0 U(B,T)/kT$ (k…Boltzmann constant) is generally $T$-dependent, in contrast to the usual analysis. If $U(B,T)$ is separated as $U = U_0(B)(1-t)^q$, $t = T/T_c$, the temperature dependent derivative $U'$ for each magnetic field reads as follows:

$$U'(T) = -d\ln R/dT^{-1} = [U_0(B)(1-t)^q - kT][1 + qt/(1-t)]. \qquad (2)$$

Whereas the exponent $q$ is field-dependent as proposed by Zhang et al. [11], it is set constant here in order to reduce arbitrariness. The green line in Fig. 2(a) is a fit to the $U'$ data at 13 T with eq. 2, and the green lines in Figs. 1(a) and 1(b) show the corresponding fits with eq. 1 to all $R(T)$ data. Best fits were achieved with a $T_c$ value of 44.0 K (corresponding well to the onset $T_c$ value) and $q$ = 1.55. The zero-temperature activation energy $U_0$ is shown in Fig. 2(c) in comparison to values from the usual linear fits to the Arrhenius plots. Both show $B^{-1}$ dependence in magnetic fields above 4 T, indicative of collective pinning. For lower magnetic fields, the logarithmic slope of $U_0(B)$ is gradually decreasing, which is expected for the transition to the single vortex pinning regime. The Arrhenius plot values (corresponding to $q$ = 1) are underestimations but may be regarded as $U_0$ values near the glass-liquid transition. $U_0$ is very comparable to literature data ($q$=1) on a NdFeAs(O,F) single crystal [12], whereas the SmFeAsO$_{0.85}$ single crystal [5] and the SmFeAs(O,F) thin film [13] show around 3 times higher values and the LaFeAs(O,F) thin film [14] one order of magnitude lower values. This is closely correlated to $T_c$ (being around 25 K (La), 45 K (Nd), and 55 K (Sm)) but individual differences in defect density and pinning strength (e.g. F-doped vs. O-deficient samples [7] and F-content dependence [15]) certainly play a role. Polycrystalline samples of Fig. 2(c)'s *Ln*1111 compounds usually exhibit slightly larger values of $U_0$ in literature (due to the averaging effect between different crystallite orientations and increased density of strongly pinning defects). At medium fields, $U_0$ is inversely proportional to $B$ for all samples in Fig. 2(c), at high fields $U_0(B)$ is well approximated by $B^{-a}(1-B/B_0)^b$ [16] with $a$=$b$=1 and reasonable values for $B_0$, which is close to the irreversibility field at 0 K.

The theory of vortex glass-liquid transition [1] predicts that resistance and current (or more generally resistivity $\rho$ and current density $J$) can be scaled according to

$$V/I \, (T-T_g)^{-v(z+2-d)} \tag{3}$$

$$I/T^\alpha \, (T-T_g)^{v(1-d)} \tag{4}$$

where v and z are the static and the dynamic critical exponent respectively and d the dimensionality of the system. Eq. 4 was modified by the factor $1/T^\alpha$ with α = [0,1] in order to check for these both scaling methods found in the literature. Yamasaki *et al.* [17] pointed out that if the scaling parameter $J_0$ (crossover current density) is proportional to k$T$ as predicted by Koch *et al.*, α should be 1, as is found for several compounds such as $YBa_2Cu_3O_7$ [18], Co-doped Ba122 [4], $SmFeAsO_{0.85}$ [5], and $MgB_2$ [19]. Many other studies (including on similar samples as listed above and also interestingly by Koch *et al.* themselves [20]) find good scaling instead for α = 0 (i.e. without the extra factor $1/T$). This can be explained by $J_0$ being proportional to a temperature-independent thermal-fluctuation activation energy $U_{tf}$ instead of k$T$ [17]. For the sample in this study, scaling was also only possible for α = 0.

The V(I) curve exactly at $T_g$ is following a perfect power law with exponent

$$n_{Tg} = (z+1)/(d-1) \tag{5}.$$

We determined this temperature by the disappearance of the quadratic term of second-order polynomial fits to the lnV(lnI) curves, adopted from Ref. [21]. Fig. 3 shows the set of V(I) curves (a), their logarithmic derivatives near $T_g$ (b), and the scaling according to eqs. 3 and 4 for d = 3 (3D case) for B = 13 T (c) exemplarily. The behavior of the other data sets is accordingly and of the same quality. The V(I) curve closest to $T_g$ not only shows the smallest variance in n value for different fit procedures but also the largest range of constant logarithmic derivative, Fig. 3(b). From scaling behavior alone, it is not possible to discern between the different cases 3D vortex glass (VG), (quasi-)2D vortex glass and Bose glass (BG, for presence of strong columnar pinning centers), since their respective scaling parameters are related to each other via:

BG-VG ($d=3$):  $v_{BG} = 2/3 \, v_{VG}$, $z_{BG} = (3z_{VG}+1)/2$ (6)

2D-3D (VG):  $v_{2D} = 2v_{3D}$, $z_{2D} = (z_{3D}-1)/2$ (7).

Certain types of glass phases have to be ruled out instead by their parameters lying outside the expected range, i.e. $v_{VG}$ = 1...2 and $z_{VG}$ > 4 and usually < 6 for vortex glass, and $v_{BG}$ ~ 0.8...1.8 and $z_{BG}$ ~ 6...9 for Bose glass [22]. For the NdFeAs(O,F) film of this study, a Bose glass could be ruled out in that way ($z \gg 9$) and also because strong correlated c-axis-aligned defects are unlikely in such MBE-grown films. Typical defects are uncorrelated regions of differing stoichiometry in the range of 10...20 nm as measured via atom probe tomography (not shown here). Discerning between 2D and 3D via the scaling parameters alone is not possible for the investigated sample (2D: v>2 and z in the expected range, 3D: v in the range but z≫6). On the other hand, the slope of $B_{c2}\|c$ near $T_c$, $dB_{c2}/dT|_{Tc}$ = -2.1 T/K, and a $T_c$ of 44.2 K result in an orbital upper critical field $B_{c2}^{orb}$(0 K) = 64 T and therefore in a coherence length $\xi_{ab}$ = 2.25 nm. With a typical anisotropy parameter γ for this type of NdFeAs(O,F) films with thickness > 50 nm of around 4.5 [23], the coherence length in c-direction can be estimated to $\xi_c$ ~ 0.5 nm. The crossover temperature between 2D and 3D behavior in layered superconductors, $T_{cr} = (1-2(\xi_c/d)^2)T_c$, where d is here the inter layer distance, i.e. the sample's c-axis parameter of 0.8543 nm, is therefore around 14 K. Even though close to the transition to 2D behavior, the measured glass-liquid temperatures of this study are within the 3D region above $T_{cr}$, which leaves 3D vortex glass scaling as best option, and the large (and T-dependent) z values have a different reason as discussed below.

As visible in Fig. 3(b), $T_g$ at 13 T is 15.2 ± 0.4 K (the maximum error in $T_g$ was estimated by the step size in T), and $n_{Tg}$ = 4.45. With eq. 5, z is 7.9 ± 0.1 at 13 T. The resistance will go to zero for $T \to T_g$ as

$$R \sim (T-T_g)^s, \text{ with exponent } s = v(z+2-d), \tag{8}$$

within the critical region (Vogel-Fulcher relationship). That means, $T_g$ and $z$ being known, $v$ can be evaluated from the behavior of $R$ near $T_g$. If eq. 8 holds, the derivative $D = (d\ln R/dT)^{-1}$ goes linearly to zero for $T \to T_g$ with slope $1/s$. This is shown in Fig. 2(b) for 13 T. The value of $s$ in that case is $8.8 \pm 0.1$, and the according value of $v = 1.27 \pm 0.05$. The two derivatives $U'$ and $D$ are related as $U' = T^2/D$. That means they can easily be converted in the other form and cross-checked for consistency (dashed lines in Figs. 2(a) and 2(b)) to reduce the errors in $U_0$ and $s$ (and hence $v$). The temperature $T^*$ ($23.7 \pm 0.2$ K at 13 T) at which both derivatives start to fail to fit the data determines the upper bound of the critical region. It is worth mentioning that solely from $R(T)$ curves $T_g$ can only be estimated with relatively large error bars in many cases, and $T_g$ and $T^*$ are usually overestimated as well as $s$ and hence $v$ underestimated.

As Liu et al. showed, an effective activation energy $U_{eff} = kT (T_c-T)/(T-T_g)$ for thermal activation of flux motion in the critical region leads to the observed power law dependence of $R$ near $T_g$ and is consistent with the vortex-glass theory [24]. That means, $R$ can be scaled in the critical region with a temperature scale $(T-T_g)/(T_c-T_g)$. Figure 1(c) compares these dependencies for the measured fields between 1 and 13 T. Due to an increasing $z$ value for increasing $T$ (or decreasing $B$), $s$ (expected values 2.7…8.5 for a 3D vortex glass) is not constant and perfect scaling is not observed. However, the critical region between $T_g$ and $T^*$ can be identified clearly. A similar scaling approach by Andersson et al. [25] with temperature scale $[T(T_c-T_g)/T_g(T_c-T)]-1$ also does not lead to scaling for this sample and, furthermore, does not resolve the critical region correctly. The latter scaling approach has nonetheless been successfully applied to Ba122 single crystals [26] and $(Li_{1-x}Fe_xOH)FeSe$ [27]. The red lines in Fig. 1(b) are fits according to $R = R_n \exp(-U_{eff}/kT)^s$ [24] with $U_{eff}$ given above.

The results of this study are summarized in Fig. 4: Three clearly distinct regions in the magnetic field-temperature phase diagram, Fig. 4(a), of this Nd1111 film are visible: The glass liquid transition ($T_g$ or $B_g(T)$) separates vortex glass and vortex liquid. The vortex liquid shows a critical region near $T_g$ as well as a thermal-activation region, which are separated by $T^*$ (i.e. $B^*(T)$). Nonetheless, the resistance within the critical region can also be described by thermal activation with an effective pinning potential $U_{eff}$ as described above. The dashed line separating thermal activation and transition region (determined by inhomogeneity broadening and fluctuations) denotes the temperatures at which eqs. 1 and 2 fail to fit $R(T)$, Fig. 1(a). Whereas the static exponent v, Fig. 4(b), is constant with temperature (or field) at around 1.35 within the error bars (regarding scaling behavior), the dynamic critical exponent z, Fig. 4(c), is increasing with temperature (or as often visualized in literature decreasing with increasing field) as mentioned above. Forcing z to be constant leads to non-scaling of $V(I)$ and offsets in $T_g$. Similarly high values of z (while v is more or less constant) have been observed for $YBa_2Cu_3O_7$ films by Voss-de Haan *et al.* [28]. The authors pointed out that reducing the range in electrical field of the $V(I)$ curves would lead to lower z values. The dashed red line is a fit according to $B_g \sim (1-T/T_{g0})^m$ with $m = 1.5 \pm 0.1$, where $T_{g0}$ is the glass-liquid transition temperature in zero applied field. An exponent m around 1.5 is often found for sufficiently high magnetic fields.

The theory of glass-liquid transition has not been unquestioned. Coppersmith *et al.* [29] showed very early that the (scaling) $V(I)$ curves could in principle be explained by pure thermal activation of flux line motion, where the counterargument, however, is the universality of the scaling parameters [1]. Similar results were achieved, e.g., by Hu *et al.* [30] for theoretical $V(I)$ curves on $YBa_2Cu_3O_7$ films. Strachan *et al.* [31] showed that often a unique glass-liquid temperature cannot be deduced from logarithmic derivatives of the $V(I)$ curves nor from successful scaling, which might explain the large range of the critical parameters in literature. They found a peak in dln$V$/dln$I$ instead of a constant value at $T_g$. Such an effect, however, was not seen for our sample, Fig. 3(b). Sullivan et al. [32] finally pointed out that the $V(I)$ curves near the transition can be heavily altered by noise effects. We therefore ensured to measure $R(T)$ as well as $V(I)$ with minimum noise and to avoid possible spurious effects as well as to disregard any data which may be affected by run-away temperature. All of the above critics come to the conclusion that scaling alone is not sufficient evidence for the presence of a vortex phase transition. Other authors on the contrary conclude that a glass-liquid transition and thermal activation of flux motion do simply not necessarily have to exclude each other [24], [25], see above. The group of Matsushita, e.g. [33], developed a model, which describes the influence of disorder and $J_c$ variance on the scaling behavior of the $V(I)$ curves, relating the critical exponents v and z with new parameters $m = (z-1)/2$ ($J_c$ variance) and $δ = 2v$ ($T$-dependence of $J_c$), which leads to, in general, $T$-dependent parameters v and z. This model can explain the large and $T$-dependent z values for our sample.

In the end, however, it is a matter of interpretation whether one regards the transport behavior as glass-liquid transition altered by thermal activation and disorder or as thermal activation mimicking a phase transition, since it is not decidable from the data alone according to literature. Whether possibly cleaner Nd1111 (and other $Ln$1111) film samples with maximized $T_c$ would show glass-liquid behavior with constant critical parameters in the expected range is an interesting task for future investigations.

**Conclusion**

In summary, the temperature dependence of the resistance, $R(T)$, as well as the voltage-current characteristics, $V(I)$, of the NdFeAs(O,F) film of this study (grown by MBE, 90 nm thickness) are well explained by the vortex glass theory with a transition to the vortex liquid at the glass-liquid transition temperature $T_g$. Whereas the static critical exponent v of around 1.35 lies within the expected range (1…2) for a vortex glass, the dynamic exponent *z* is considerably larger than the expected upper bound of 6 and more importantly temperature-dependent, i.e. non-universal. This can be explained by thermally activated flux motion in the liquid region (also within the critical region $T_g$…$T^*$) in combination with $T_c$ and hence $J_c$ variation in the sample.

**Acknowledgment**

The authors thank Rainer Nast (KIT) for laser-structuring of the sample and A. Gruber (KIT) for technical support. This work was partially supported by the JSPS Grant-in-Aid for Scientific Research (B) Grant Number 16H04646.

**Figure Captions**

FIG. 1. Temperature dependence of the resistance, $R(T)$, for self-field as well as between 1 T and 13 T (step size 2 T) in (a) normal and (b) Arrhenius representation. The green lines are fits according to eq. 1 and the red lines according to $R = R_n \exp(-U_{eff}/kT)^s$. (c) Scaling behavior of the resistance curves at fields between 1 and 13 T (step size 2 T) according to ref. [24]. The data of the critical region show power law behavior with exponent $s = v(z-1)$ changing from 8.8 to 10.5 due to the $T$-dependence of $z$, Fig. 4(c). Consequently, true scaling of all curves is not observed.

FIG. 2. Derivatives of $R(T)$ at $B = 13$ T. (a) Apparent activation energy $U'(T)$, (b) derivative $D = (d\ln R/d\ln T)^{-1}$. The green lines are fits according to eq. 2, and the red lines according to eq. 8. Both derivatives are related as $U' = T^2/D$. (c) Magnetic field dependence of the thermal activation energy for flux motion $U_0$ for $B\|c$ of the NdFeAs(O,F) film (red) in comparison to NdFeAs(O,F) [12] and SmFeAsO$_{0.85}$ [5] single crystals as well as SmFeAs(O,F) [13] and LaFeAs(O,F) [14] epitaxial thin film. Open symbols: films, closed symbols: single crystals. $q=1$: from linear fits of the Arrhenius plots (with $T_c = 44.2$ K), $q=1.55$: determined via fits of $U'$ with eq. 2 (with $T_c = 44.0$ K). The former values can be regarded as apparent activation energies near the glass-liquid transition. At medium fields, $U_0$ is inversely proportional to $B$, at high fields $U_0(B)$ is well approximated by $B^{-a}(1-B/B_0)^b$ with $a=b=1$ and reasonable values for $B_0$. The errors in $U_0$ are roughly the symbol size.

FIG. 3. (a) $V(I)$ curves, (b) their logarithmic derivatives and (c) their scaling near $T_g$ of the investigated NdFeAs(O,F) film exemplarily for $B = 13$ T.

FIG. 4. (a) *B-T* phase diagram for $B \parallel c$ of a 90 nm thick NdFeAs(O,F) film on MgO(001) together with the temperature dependencies of the critical exponents (b) v and (c) z.

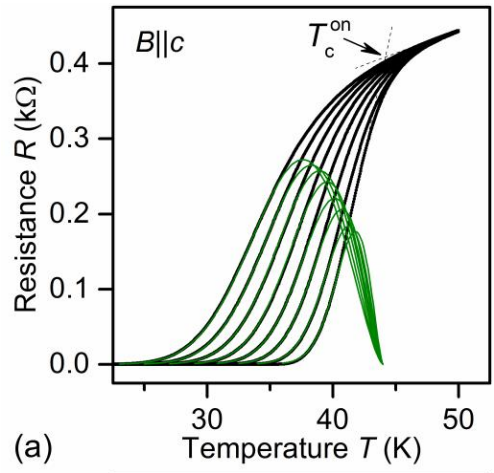
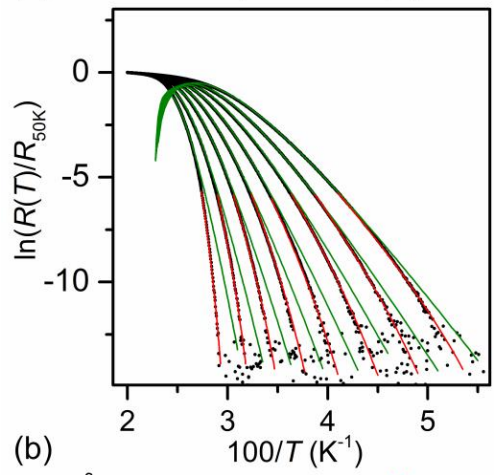
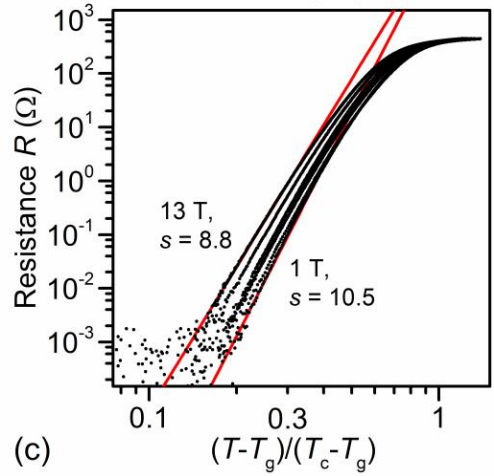

Figure 1

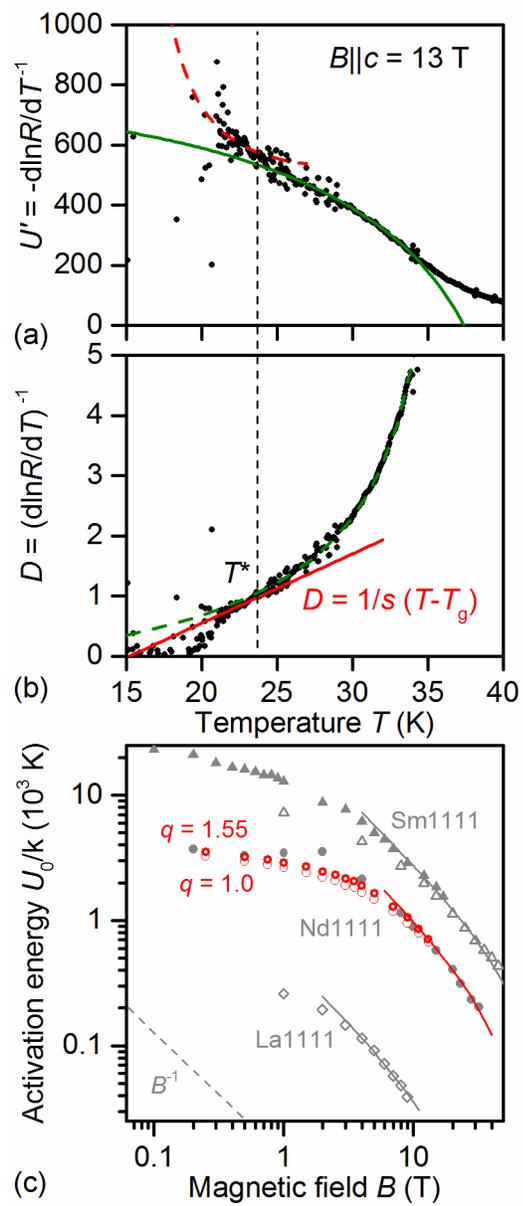

Figure 2

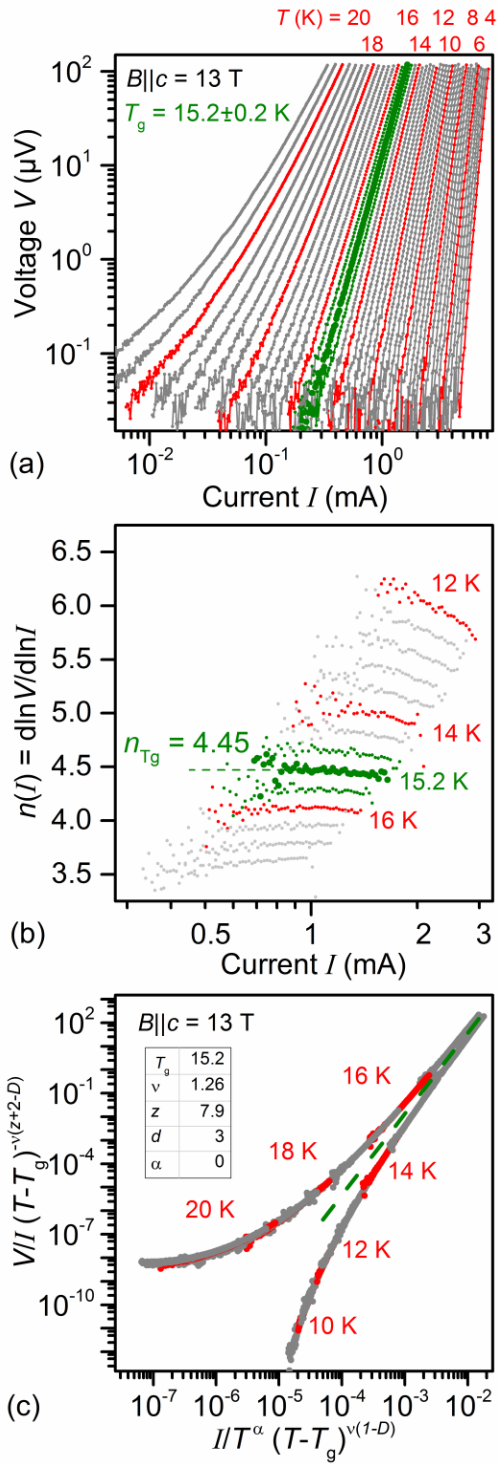

Figure 3

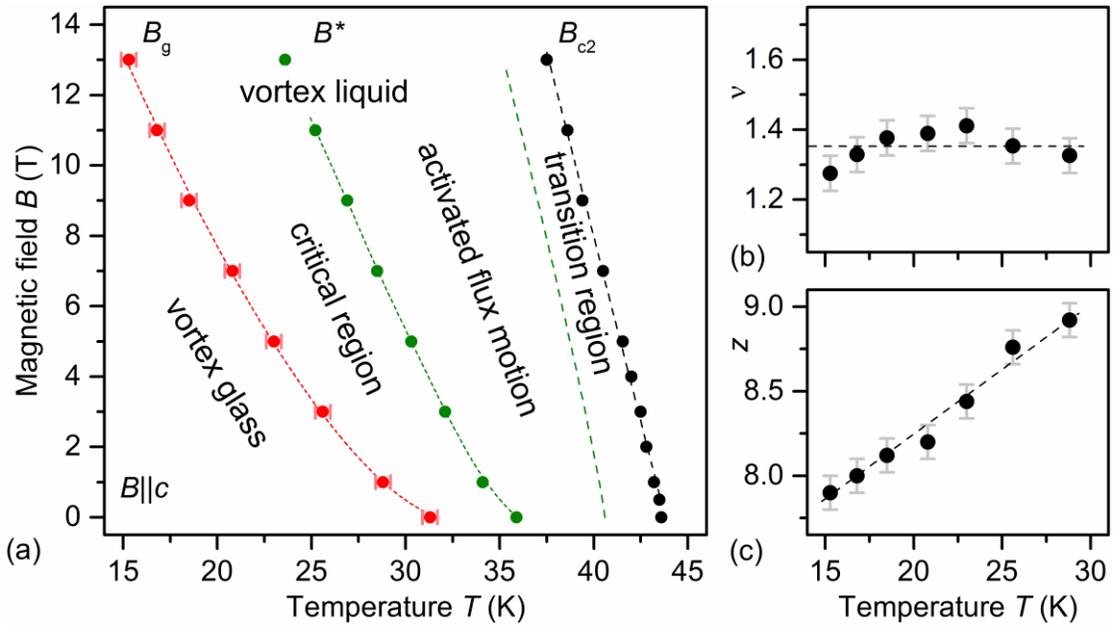

Figure 4